\def\be{\begin{equation}}
\def\ee{\end{equation}}
\def\ba{\begin{array}}
\def\ea{\end{array}}
\def\

\def\l{\lambda}

\makeatletter

\newcommand{\Rmnum}[1]{\expandafter\@slowromancap\romannumeral #1@}
\makeatother

\documentclass[prl,showpacs,a4paper,amsmath,amssymb]{revtex4}
\usepackage{amsfonts}
\usepackage{epsfig,graphicx}
\usepackage{epstopdf}
\usepackage{subfigure}
\usepackage{amsmath}
\usepackage{lipsum}
\usepackage{amscd,color}
\input amssym.def
\newtheorem{theorem}{Theorem}
\newtheorem{lemma}{Lemma}
\newtheorem{proposition}{Proposition}
\begin{document}
\title{The reduction of the number of incoherent Kraus operations for qutrit systems}
\author{Jiahuan Qiao$^1$}
\author{Lingyun Sun$^1$}
\author{Jing Wang$^1$}
\author{Ming Li$^{1}$}
\author{Shuqian Shen$^1$}
\author{Lei Li$^1$}
\author{Shaoming Fei$^{2,3}$}
\affiliation{$^1$College of the Science, China University of
Petroleum, 266580 Qingdao, China\\
$^2$ School of Mathematical Sciences, Capital Normal University,
 100048 Beijing, China\\
$^3$ Max-Planck-Institute for Mathematics in the Sciences, 04103
Leipzig, Germany}
\begin{abstract}

Quantum coherence is a fundamental property that can emerge within
any quantum system. Incoherent operations, defined in terms of the
Kraus decomposition, take an important role in state transformation.
The maximum number of incoherent Kraus operators has been presented in
[A. Streltsov, S. Rana, P. Boes, J. Eisert, Phys. Rev. Lett. 119.
140402 (2017)]. In this work, we show that the number of incoherent
Kraus operators for a single qubit can be reduced from 5 to 4 by
constructing a proper unitary matrix. For qutrit systems we further
obtain 32 incoherent Kraus operators, while the upper bound in the
research of Sterltsov gives 39 Kraus operators. Besides, we reduce
the number of strictly incoherent Kraus operators from more than 15
to 13. And we consider the state transformation problem for these two types of operations in single qutrit systems.

\end{abstract}
\smallskip

\pacs{03.67.-a, 02.20.Hj, 03.65.-w} \maketitle
\section{\Rmnum{1}.Introduction}
Quantum resource theories
\cite{PhysRevLett.115.070503,RevModPhys.91.025001} offer a powerful
framework for understanding the natural change of certain physical
properties in a physical system and their applications for quantum
technology. In recent years, a lot of works on the development of
quantum resource theory in different physics fields have been done,
such as the quantum resource theory of entanglement
\cite{RevModPhys.81.865}, the quantum resource theory of
thermodynamics \cite{PhysRevLett.111.250404}, the quantum resource
theory of coherence \cite{PhysRevLett.113.140401} and so on. The
general structure of quantum resource theory has three ingredients
in common£ºfree states, free operations
 and resource states. The basic requirement of resource theory
is that free operations cannot generate a resource state from a free one. Free states can be created and performed
at no cost, and any state outside of the set of free states is called resource state. For example, in entanglement
theory, free states are separable states, and free operations are local operations and classical communication. While in the resource theory of coherence, free states are incoherent states, and free operations are incoherent operations. Quantum resources can be states with quantum correlations\cite{Gyongyosi1,Gyongyosi2}.

As an important physical resource, quantum
coherence \cite{humingliang,xi,Yang2018,tao} has been used in a variety of physical
tasks in quantum information processing, such as quantum algorithm
\cite{Hillery2016Coherence}, quantum thermodynamics
\cite{Lostaglio2011Description,Lostaglio2015Thermodynamic},
metrology \cite{Micadei2015Coherent}, and quantum biology
\cite{Lloyd2011Quantum}. Let $\{|i\rangle\}$ $(i=1,...,d)$ be a
paticular basis in a $d-$dimensional Hilbert space $\mathcal{H}_d$.
A state is called incoherent state if it is diagonal in this basis
and otherwise
 coherent. The structure of the incoherent states is as follows
\begin{equation}\label{incoherentstate}
\delta=\sum_{i=1}^d\delta_i|i\rangle\langle i|,
\end{equation}
where $\sum_{i=1}^d\delta_i=1$.

Depending on the different physical requirement, there exist different types of incoherent operations.
The important free operations are known as incoherent operations(IO)
 \cite{PhysRevLett.113.140401} and strictly incoherent operations(SIO) \cite{PhysRevLett.116.120404}.
We denote $\mathcal{I}$ as the set of all incoherent states. A
completely positive and trace-preserving map(CPTP) $\Phi$ is said to
be an IO if $\Phi$ has a Kraus operator representation $\{K_n\}$
such that $K_n\rho K_n^\dagger/Tr[K_n\rho K_n^\dagger]\in\mathcal
{I}$ for all $n$ and $\rho\in\mathcal{I}$, while SIO require further
$\{K_n\}$ and $\{K_n^\dagger\}$ are incoherent.

Recently, A. Streltsov et al. in \cite{PhysRevLett.119.140402} have
derived the upper bound of the number of incoherent Kraus operators
in a general incoherent operation. For any single qubit IO, the
canonical representation of the Kraus operator is given by the set
\begin{equation}\label{5Krausoperator}
\begin{split}
\left\{
\left(\begin{array}{ccc}
       a_1&b_1\\0&0\\
       \end{array} \right),
\left(\begin{array}{ccc}
       0&0\\a_2&b_2\\
       \end{array} \right),
\left(\begin{array}{ccc}
        a_3&0\\0&b_3\\
       \end{array} \right),
\left(\begin{array}{ccc}
       0&b_4\\a_4&0\\
       \end{array} \right),
\left(\begin{array}{ccc}
       a_5&0\\0&0\\
       \end{array} \right)
\right\},
\end{split}
\end{equation}
where $a_i\in\mathbb{R}$, $b_j\in\mathbb{C}$. Moreover, $a_i$ and
$b_j$ should satisfy the equalities
$\sum_{i=1}^5a_i^2=\sum_{j=1}^4|b_j|^2=1$ and $a_1b_1+a_2b_2=0$.
Later, some scholars have reduced the optimal number of incoherent
Kraus operators on qubit systems to 4 by using the Choi-Jamio${\l}$kowski-Sudarshan
 matrix \cite{Rana_2018}, which is proved to be
optimal. In this work, we reduce the number of qubit and qutrit
incoherent Kraus operators by constructing proper unitary matrices.
We show that the number of incoherent Kraus operators for a single
qubit can be reduced from 5 to 4. For qutrit systems we obtain 32
incoherent Kraus operators, while the upper bound in the research of
Sterltsov gives 39 Kraus operators. Besides, we reduce the number of
strictly incoherent Kraus operators from more than 15 to 13.
 Lastly, we consider the state transformation via SIO and
IO in qutrit system. And we find the achievable region about the set of final
states from a given initial qutrit state by all possible qutrit IOs.

\section{\Rmnum{2}.The upper bound of (strictly) incoherent operators for qutrit system}

Recently, the structure of incoherent and strictly incoherent
operations is studied in \cite{PhysRevLett.119.140402,Rana_2018}. As
mentioned in \cite{PhysRevLett.119.140402}, any single qubit IO can
be decomposed into 5 incoherent Kraus operators using the structure
of IO. Similarly, the number of incoherent Kraus operators can be
reduced to 39 for any single qutrit incoherent operation. Besides,
the upper bound of the number of strictly incoherent operator is less
than 15. In the following, we first introduce an isometry
 about the two sets of Kraus decompositions which give rise to the same quantum operation.
\begin{lemma}
The two sets of Kraus operators $\{K_j\}$ and $\{L_i\}$ are Kraus
decompositions of the same quantum operation if and only if there is
a unitary matrix $U$ such that \cite{Nielsen2010Quantumbibtex}
\begin{equation}
  L_i=\sum_{j}U_{i,j}K_j.
\end{equation}
\end{lemma}

Therefore, according to the above result, the number of Kraus operators of a
quantum operation is finite.
There must be a set with the least number of
Kraus operators. Firstly, let's study the qubit case. By using the
properties of Lemma 1 and the qubit incoherent Kraus operator, we
find the following conclusion.

\begin{proposition}
Every qubit IO can be decomposed into four incoherent Kraus operators. The canonical representation
of the Kraus operators is given by the set
\begin{equation}\label{4Krausoperator}
\begin{split}
\left\{
\left(\begin{array}{ccc}
       a_1&b_1\\0&0\\
       \end{array} \right),
\left(\begin{array}{ccc}
       0&0\\a_2&b_2\\
       \end{array} \right),
\left(\begin{array}{ccc}
        a_3&0\\0&b_3\\
       \end{array} \right),
\left(\begin{array}{ccc}
       0&b_4\\a_4&0\\
       \end{array} \right)
\right\},
\end{split}
\end{equation}
where $a_i\in\mathbb{R}$, $b_j\in\mathbb{C}$ satisfing the
equalities $\sum_{i=1}^4a_i^2=\sum_{j=1}^4|b_j|^2=1$ and
$a_1b_1+a_2b_2=0$.
\end{proposition}

See section A in the supplementary material for the proof of the proposition.

Using the Choi-Jamio${\l}$kowski-Sudarshan matrix for a quantum
operation, Rana et al. have proved that the optimal number of
incoherent Kraus operators for an incoherent qubit operation is
four\cite{Rana_2018}. However, we observe that it is more convenient to draw the
conclusion using the isometry of Kraus decompositions. For most
incoherent operations, the above result is the optimal form of
incoherent Kraus decomposition. We cannot find a general unitary
matrix to reduce the number of incoherent Kraus
 operator. But some special quantum operations could be decomposed into
least four incoherent Kraus operators, such as the phase damping
channel and amplitude damping channel \cite{Yang2018}.

For qutrit system, any incoherent operation admits a decomposition
with at most $39$ incoherent Kraus operators. A canonical
representation of the Kraus operators for a qutrit IO can be
obtained from the proof of Proposition 5 in Ref.
\cite{PhysRevLett.119.140402} as follows,

 \begin{equation}
\begin{split}
  &K_1=\left[\begin{array}{ccc}
       a_1&b_1&c_1\\0&0&0\\  0&0&0\\
       \end{array} \right],
 \quad K_2= \left[ \begin{array}{ccc}
       a_2&b_2&0\\0&0&c_2\\  0&0&0\\
       \end{array}\right],
 \quad K_3= \left[ \begin{array}{ccc}
       a_3&b_3&0\\0&0&0\\ 0&0&c_3\\
       \end{array} \right],
 \quad K_4= \left[ \begin{array}{ccc}
       a_4&b_4&0\\ 0&0&0\\ 0&0&0\\
       \end{array}\right],
 \quad K_5= \left[\begin{array}{ccc}
       0&0&c_5\\a_5&b_5&0\\0&0&0\\
       \end{array}\right],\\
 &K_6= \left[ \begin{array}{ccc}
       0&0&0\\ a_6&b_6&c_6\\ 0&0&0\\
       \end{array} \right],
\quad K_7= \left[ \begin{array}{ccc}
        0&0&0\\ a_7&b_7&0\\ 0&0&c_7\\
        \end{array} \right],
\quad K_8= \left[ \begin{array}{ccc}
        0&0&0\\ a_8&b_8&0\\ 0&0&0\\
       \end{array} \right],
\quad K_9= \left[ \begin{array}{ccc}
       a_9&0&c_9\\ 0&b_9&0\\0&0&0\\
      \end{array} \right],
\quad K_{10}= \left[ \begin{array}{ccc}
       a_{10}&0&0\\ 0&b_{10}&c_{10}\\0&0&0\\
      \end{array} \right],\\
  &K_{11}= \left[ \begin{array}{ccc}
       a_{11}&0&0\\ 0&b_{11}&0\\0&0& c_{11}\\
      \end{array} \right],
\quad  K_{12}= \left[ \begin{array}{ccc}
       a_{12}&0&0\\ 0&b_{12}&0\\0&0&0\\
      \end{array} \right],
\quad  K_{13}= \left[ \begin{array}{ccc}
        0&b_{13}&c_{13}\\ a_{13}&0&0\\0&0&0\\
      \end{array} \right],
\quad  K_{14}= \left[ \begin{array}{ccc}
        0&b_{14}&0\\ a_{14}&0&c_{14}\\0&0&0\\
      \end{array} \right],\\
   &K_{15}= \left[ \begin{array}{ccc}
        0&b_{15}&0\\ a_{15}&0&0\\0&0&c_{15}\\
      \end{array} \right], 
\quad  K_{16}= \left[ \begin{array}{ccc}
        0&b_{16}&0\\ a_{16}&0&0\\0&0&0\\
      \end{array} \right],
\quad  K_{17}= \left[ \begin{array}{ccc}
       a_{17}&0&c_{17}\\0&0&0\\ 0&b_{17}&0\\
      \end{array} \right],
\quad  K_{18}= \left[ \begin{array}{ccc}
       a_{18}&0&0\\0&0&c_{18}\\ 0&b_{18}&0\\
      \end{array} \right],\\
   &K_{19}= \left[ \begin{array}{ccc}
       a_{19}&0&0\\0&0&0\\ 0&b_{19}&c_{19}\\
      \end{array} \right],
\quad  K_{20}= \left[ \begin{array}{ccc}
       a_{20}&0&0\\0&0&0\\ 0&b_{20}&0\\
      \end{array} \right], 
\quad  K_{21}= \left[ \begin{array}{ccc}
       0&0&c_{21}\\a_{21}&0&0\\ 0&b_{21}&0\\
      \end{array} \right],
\quad  K_{22}= \left[ \begin{array}{ccc}
       0&0&0\\a_{22}&0&c_{22}\\ 0&b_{22}&0\\
      \end{array} \right],\\
   &K_{23}= \left[ \begin{array}{ccc}
       0&0&0\\a_{23}&0&0\\ 0&b_{23}&c_{23}\\
      \end{array} \right],
\quad  K_{24}= \left[ \begin{array}{ccc}
       0&0&0\\a_{24}&0&0\\ 0&b_{24}&0\\
      \end{array} \right],
\quad  K_{25}= \left[ \begin{array}{ccc}
       0&0&c_{25}\\ 0&b_{25}&0\\a_{25}&0&0\\
      \end{array} \right],
\quad  K_{26}= \left[ \begin{array}{ccc}
       0&0&0\\ 0&b_{26}&c_{26}\\a_{26}&0&0\\
      \end{array} \right],\\
    &K_{27}= \left[ \begin{array}{ccc}
       0&0&0\\ 0&b_{27}&0\\a_{27}&0& c_{27}\\
      \end{array} \right],
\quad  K_{28}= \left[ \begin{array}{ccc}
       0&0&0\\ 0&b_{28}&0\\a_{28}&0&0\\
      \end{array} \right],
\quad  K_{29}= \left[ \begin{array}{ccc}
       0&0&c_{29}\\ 0&0&0\\a_{29}&b_{29}&0\\
      \end{array} \right],
\quad  K_{30}= \left[ \begin{array}{ccc}
       0&0&0 \\ 0&0&c_{30}\\a_{30}&b_{30}&0\\
      \end{array} \right],\\
   &K_{31}= \left[ \begin{array}{ccc}
       0&0&0 \\ 0&0&0\\a_{31}&b_{31}&c_{31}\\
      \end{array} \right],
\quad  K_{32}= \left[ \begin{array}{ccc}
       0&0&0 \\ 0&0&0\\a_{32}&b_{32}&0\\
      \end{array} \right],
\quad  K_{33}= \left[ \begin{array}{ccc}
        0&b_{33}&c_{33}\\0&0&0 \\ a_{33}&0&0\\
      \end{array} \right],
\quad  K_{34}= \left[ \begin{array}{ccc}
        0&b_{34}&0\\0&0& c_{34} \\ a_{34}&0&0\\
      \end{array} \right],\\
   &K_{35}= \left[ \begin{array}{ccc}
        0&b_{35}&0\\0&0&0 \\ a_{35}&0& c_{35}\\
      \end{array} \right],
\quad  K_{36}= \left[ \begin{array}{ccc}
        0&b_{36}&0\\0&0&0 \\ a_{36}&0&0\\
      \end{array} \right],
\quad  K_{37}= \left[ \begin{array}{ccc}
        0&0&0 \\0&0&0 \\ a_{37}&0&0\\
      \end{array} \right],
\quad  K_{38}= \left[ \begin{array}{ccc}
       a_{38}&0&0\\ 0&0&0 \\0&0&0 \\
      \end{array} \right],
\quad  K_{39}= \left[ \begin{array}{ccc}
     0&0&0 \\ a_{39}&0&0\\0&0&0 \\
      \end{array} \right].
\end{split}
\end{equation}
Note that the second-order submatrix of the above Kraus operators have similar matrix forms to the incoherent Kraus operators in the qubit system. Based on this relationship, we can draw the following conclusions.
\begin{theorem}
Any incoherent operation acting on a single qutrit system admits a decomposition with at most 32 incoherent Kraus operators.
\end{theorem}

See section B in the supplementary material for the proof of the theorem.

\begin{theorem}
 Any strictly incoherent operation acting on a single qutrit system admits a decomposition with at most 13 strictly incoherent Kraus
operators.
\end{theorem}

See section C in the supplementary material for the proof of the theorem.

In 3-dimension Hilbert space, an arbitrary quantum state is expressed as
\begin{equation}
  \rho=\frac{1}{3}I+\frac{1}{2}\sum_{i=1}^{8}t_i\lambda_i,
\end{equation}
where $\vec{t}=\{t_1, t_2,...,t_{8}\}$ is the $8$ dimensional Bloch
vector, $\lambda_i$ is a generator of SU(3), where the length of $\vec{t}$
should be less than or equal to $\frac{2}{\sqrt{3}}$
\cite{KIMURA2003339}. In order to visualize the state transformation
via single qutrit SIO and IO, we consider two-dimensional sections
of $\sum_3(i,j)$ \cite{JAKOBCZYK2001383} which are constructed as
$\sum_3(i,j)=\{\textbf{t}\in B(\mathbb{R}^8):
\textbf{t}=\{0,...,0,t_i,0,...,t_j,...,0\} \}$. For a given Bloch
vector of
 $\textbf{t}=\{{0,...,0,t_i,0,...,t_j,...,0} \}$, we can find the achievable region for the final state
 $\textbf{m}=\{{0,...,0,m_i,0,...,m_j,...,0} \}$ via single qutrit SIO and IO.
  In the two-dimensional sections, we can find the limited conditions, the proof can be found in section D of the supplementary material.

1: In the $\{m_1,...,m_6\}-\{m_7, m_8\} $ plane, the following
inequalities should be satisfied:
\begin{equation}
\begin{split}
  m_{i}^2&\leq  t_{i}^2, \quad \{i=1,...,6\}\\
  m_{7}&\in[\frac{1-\sqrt{3}}{3}, \frac{2}{\sqrt{3}}],\\
  m_{8}&\in[-\frac{2\sqrt{3}}{3}, \frac{2\sqrt{3}}{3}].\\
  \end{split}
\end{equation}

2: In the $\{m_7\}-\{ m_8\} $ plane, the following equality should
be satisfied:
\begin{equation}
-\sqrt{3}m_7+m_8-\frac{2\sqrt{3}}{3}=0.
\end{equation}

3: In the $\{m_1\}-\{ m_4\} $, $\{m_2\}-\{ m_5\} $ and $\{m_3\}-\{
m_6\} $ planes, the following inequalities should be satisfied
respectively:
\begin{equation}
\begin{split}
  m_{1}^2+m_{4}^2&\leq  t_{1}^2+t_{4}^2,\\
  m_{2}^2+m_{5}^2&\leq  t_{2}^2+t_{5}^2,\\
  m_{3}^2+m_{6}^2&\leq  t_{3}^2+t_{6}^2.
  \end{split}
\end{equation}

4: In the other planes, we find the following inequality:
\begin{equation}
  (|m_i|+|m_j|)^2\leq (|t_i|+|t_j|)^2.
\end{equation}

In the following Fig.\ref{figa}$-$Fig.\ref{figd}, we show the projection of the achievable region into the $\{m_1,...,m_6\}-\{m_7, m_8\} ,\{m_7\}-\{ m_8\},\{m_1\}-\{ m_4\}, \{m_2\}-\{ m_5\}, \{m_3\}-\{m_6\} $ and other planes for the corresponding different initial states. The numerical simulations of the final states in the following pictures coincide with our conclusions.
\begin{widetext}

\begin{figure}[htbp]
 \centering
\includegraphics[width=0.9\textwidth]{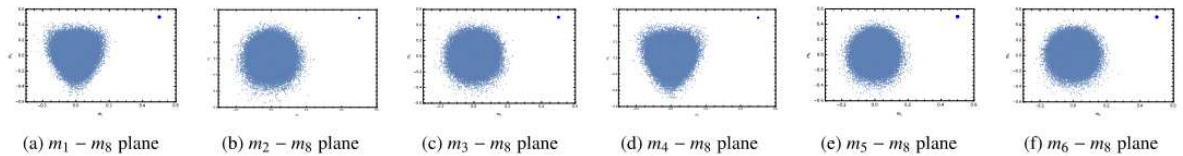}\\
\caption{The achievable region for single qutrit SIO and IO in
condition 1. The blue colored area shows the projection of the
achievable region in the $m_i-m_8$ $(i=1,...,6)$
 plane. We have set $t_i=t_8=0.5$ in the initial state.}
 \label{figa}
\end{figure}

\begin{figure}[htbp]
 \centering
\includegraphics[width=0.9\textwidth]{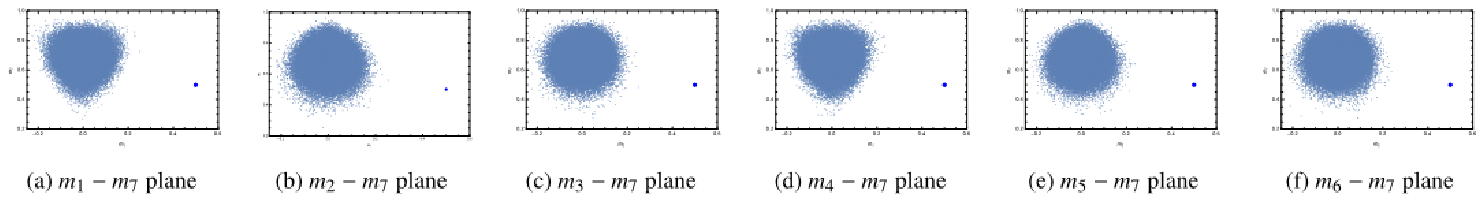}\\
\caption{The achievable region for single qutrit SIO and IO in
condition 2. The blue colored area shows the projection of the
achievable region in the $m_i-m_7$ $(i=1,...,6)$ plane. In the
initial state we set $t_i=0.5,\ t_7=0.5$ respectively.}
\label{figb}
\end{figure}

\begin{figure}[htbp]
 \centering
\includegraphics[width=0.7\textwidth]{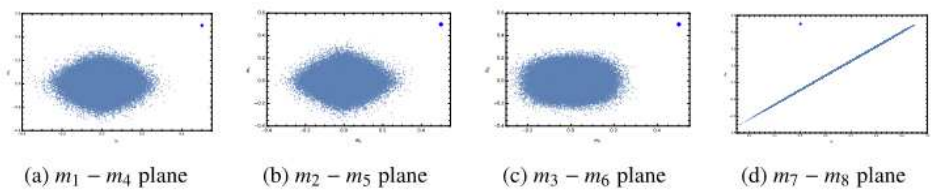}\\
\caption{The achievable region for single qutrit SIO and IO in
condition 3. The blue colored area shows the projection of the
achievable region in the $m_1-m_4$, $m_2-m_5$, $m_3-m_6$ and
$m_7-m_8$  plane, where in the initial state we set $t_i=0.5, t_j=0.5
$  respectively.}
\label{figc}
\end{figure}

\begin{figure}[!h]
\centering
\includegraphics[width=0.9\textwidth]{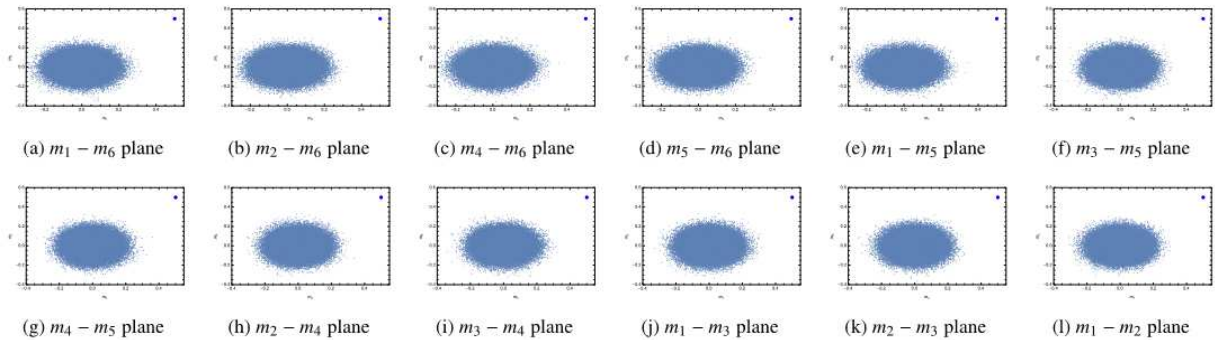}\\
\caption{The achievable region for single qutrit SIO and IO in
condition 4. The blue colored area shows the projection of the
achievable region in the planes that are not mentioned above,
where in the initial state we set $t_i=0.5, t_j=0.5$
respectively.}
\label{figd}
\end{figure}
\end{widetext}

\section{\Rmnum{3}. DISCUSSION AND Conclusion}

In this paper, we have discussed how to reduce the number of
incoherent Kraus operators. Furthermore, we have shown that the
number of incoherent Kraus operators for a single qubit can be
reduced from 5 to 4. For qutrit system, we have found that any
incoherent operation or strictly incoherent operation admits
decomposition with at most 32 or 13 Kraus operator respectively. We
have also investigated the achievable region for a fixed state via
single qutrit SIO and IO. An open question is that whether the upper
bound can be further reduced to a much tight level. Besides, it is
still yet to be solved to compute the optimal number of incoherent
Kraus operator when $d\geq 4$. We suspect that the number of d
dimensional incoherent Kraus operators is related to the number of
d-1 dimensional incoherent Kraus operators. In addition, the form of
incoherent Kraus operator in d dimension is also related to the d-1
dimensional incoherent Kraus operators. More importantly, according to the relationship
 between the superposition-free operation and incoherent operation, we can obtain the
structure of the resource theory of superposition\cite{PhysRevLett.119.230401}.

\bigskip
\noindent{\bf Acknowledgments}\, \, This work is supported by NSFC (11775306, 11701568,
11675113), the Fundamental Research Funds for the Central
Universities (18CX02035A, 18CX02023A, 19CX02050A),
Beijing Municipal Commission of Education under Grant No.
KZ201810028042, and Beijing Natural Science Foundation
(Z190005).


\newpage

\setcounter{equation}{0}
\renewcommand{\theequation}{S\arabic{equation}}

\textbf{Supplemental material for ``The reduction of the number of incoherent Kraus operations for qutrit systems''}

\section{A: The proof of proposition 1.}
Denote the incoherent Kraus operations in Eq.(\ref{5Krausoperator}) as follows
\begin{equation}
  \begin{split}
K_1&=
\left(\begin{array}{ccc}
       a_1&b_1\\0&0\\
       \end{array} \right),
\quad K_2=
\left(\begin{array}{ccc}
       0&0\\a_2&b_2\\
       \end{array} \right),
\quad K_3=
\left(\begin{array}{ccc}
       a_3&0\\0&0\\
       \end{array} \right),
\quad K_4=
\left(\begin{array}{ccc}
       0&b_4\\ a_4&0\\
       \end{array} \right),
\quad K_5=
\left(\begin{array}{ccc}
        \ast&0\\0&\ast\\
       \end{array} \right),
 \end{split}
\end{equation}
where $a_i\in\mathbb{R}$, $b_j\in\mathbb{C}$, $\ast$ denotes some
complex number. Now, select the following $4\times 4$ unitary matrix
\begin{equation}
\begin{split}
 U=\left(\begin{array}{cccc}
       ka_1&0&ka_3&0\\
       -la_3^2a_4|b_1|^2|b_4|^2&la_2b_1b_4^*(a_1^2|b_4|^2+a_3^2|b_1|^2+a_3^2|b_4|^2)
       &la_1a_3a_4|b_1|^2|b_4|^2&la_3^2a_4|b_1|^3|b_4|\\
       -ma_2a_3b_1^*b_4&-ma_3a_4|b_1|^2&ma_1a_2b_1^*b_4&ma_2a_3|b_1|^2\\
       na_3^2b_1^*b_4&0&-na_1a_3b_1^*b_4&n(a_1^2+a_3^2)|b_4|^2\\
       \end{array} \right),
 \end{split}
\end{equation}
where the parameters $k$, $l$, $m$ and $n$ are chosen as
\begin{equation}
  \begin{split}
  k^2&=\frac{1}{a_1^2+a_3^2},
  l^2=\frac{1}{a_3^4a_4^2|b_1|^4|b_4|^4+a_2^2|b_1|^2|b_4|^2(a_1^2|b_4|^2+a_3^2|b_1|^2
  +a_3^2|b_4|^2)^2+a_1^2a_3^2a_4^2|b_1|^4|b_4|^4+a_3^4a_4^2|b_1|^6|b_4|^2},\\
  m^2&=\frac{1}{a_2^2a_3^2|b_1|^2|b_4|^2+a_3^4a_4^2|b_1|^4+a_1^2a_2^2|b_1|^2|b_4|^2
  +a_2^2a_3^2|b_1|^4},
  n^2=\frac{1}{a_3^4|b_1|^2|b_4|^2+a_1^2a_3^2|b_1|^2|b_4|^2+(a_1^2+a_3^2)^2|b_4|^4}.
  \end{split}
\end{equation}

We then introduce a unitary matrix $V$ defined by
\begin{equation}\label{unitary}
 V=U\oplus I_1,
\end{equation}
where $I_1$ is the identity operator with dimension 1. According to
Lemma 1, we have
\begin{eqnarray}
L_i=
\begin{cases}
\sum_{j=1}^4V_{i,j}K_j      & {for\ 1\leq i \leq 4,} \\
K_i                         & {for\ i=5.}
\end{cases}
\end{eqnarray}
Then one computes that
\begin{equation}
\begin{split}
L_1&=
\left(\begin{array}{ccc}
       \ast &\ast\\0&0\\
       \end{array} \right),
\quad L_2=
\left(\begin{array}{ccc}
       0&0\\ \ast &\ast\\
       \end{array} \right),
\quad L_3=
\left(\begin{array}{ccc}
        \ast&0\\0&\ast\\
       \end{array} \right),
\quad L_4=
\left(\begin{array}{ccc}
       0&\ast\\ \ast&0\\
       \end{array} \right),
\quad L_5=
\left(\begin{array}{ccc}
       \ast&0\\0&\ast\\
       \end{array} \right),
\end{split}
\end{equation}
where $\ast$ denote some complex numbers that are the combinations
of $a_i$ and $b_j$. It is obvious that $L_5$ has the same form as
$L_3$. Thus, we can reduce the set $\{L_3,L_5\}$ to one Kraus
operator. This proves that every IO in qubit system can be
decomposed into at most four incoherent Kraus operators as given in
Eq.(\ref{4Krausoperator}). From the normalization property
$\sum_{i=1}^4K_i^\dagger K_i=I$, we have
$\sum_{i=1}^4a_i^2=\sum_{j=1}^4|b_j|^2=1$ and $a_1b_1+a_2b_2=0$.

\section{B: The proof of Theorem 1.}
Firstly, we choose the Kraus operators $K_{32}$, $K_{24}$, $K_{37}$
and $K_{39}$. Denote $K_{32}$, $K_{24}$, $K_{37}$ and $K_{39}$ as
$M_{1}$, $M_{2}$, $M_{3}$ and $M_{4}$ respectively. Define a
$3\times 3$ unitary matrix $U_1$ by
\begin{equation}
\begin{split}
 U_1=\left(\begin{array}{ccc}
       l_1a_{32}^\ast&0&l_1a_{37}^\ast\\
       m_1b_{32}^\ast|a_{37}|^2&m_1(|a_{32}|^2+|a_{37}|^2)b_{24}^\ast&-m_1a_{37}^\ast b_{32}^\ast a_{32}\\
       n_1a_{37}b_{24}&-n_1a_{37}b_{32}&-n_1a_{32}b_{24}
       \end{array} \right),
 \end{split}
\end{equation}
 where the parameters $l_1$, $m_1$ and $n_1$ are chosen as
 \begin{equation}
  \begin{split}
  l_1^2&=\frac{1}{a_{32}^2+a_{37}^2},\\
  m_1^2&=\frac{1}{(a_{32}^2+a_{37}^2)(|a_{37}|^2(|b_{32}|^2+|b_{24}|^2)+|a_{37}b_{24}|^2)},\\
  n_1^2&=\frac{1}{|a_{37}|^2(|b_{32}|^2+|b_{24}|^2)+|a_{37}b_{24}|^2}.
  \end{split}
\end{equation}
According to Lemma 1 and the construction of the unitary matrix such
as Eq.(\ref{unitary}), we find that the operator $L_{4}$ has the
same form as $L_{3}$. In other words,
 the set $\{M_{1},\ M_{2},\ M_{3},\ M_{4}\}$ can be reduced to the set $\{M_{1},\ M_{2},\ M_{4}\}$,
that is, the number of the incoherent Kraus operators  can be reduced
to 38.

Similarly, define $U_2$ as a unitary matrix by,
\begin{equation}
\begin{split}
 U_2=\left(\begin{array}{ccc}
       l_2a_{8}^\ast&0&l_2a_{39}^\ast\\
       m_2b_{8}^\ast|a_{39}|^2&m_2(|a_{8}|^2+|a_{39}|^2)b_{12}^\ast&-m_2a_{39}^\ast b_{8}^\ast a_{8}\\
       n_2a_{39}b_{12}&-n_2a_{39}b_{8}&-n_2a_{8}b_{12}
       \end{array} \right),
 \end{split}
\end{equation}
 where the parameters $l_2$, $m_2$ and $n_2$ are chosen as
 \begin{equation}
  \begin{split}
  l_2^2&=\frac{1}{a_{8}^2+a_{39}^2},\\
  m_2^2&=\frac{1}{(a_{8}^2+a_{39}^2)(|a_{39}|^2(|b_{8}|^2+|b_{12}|^2)+|a_{39}b_{12}|^2)},\\
  n_2^2&=\frac{1}{|a_{39}|^2(|b_{8}|^2+|b_{12}|^2)+|a_{39}b_{12}|^2}.
  \end{split}
\end{equation}
One finds the set $\{K_{8},\ K_{12},\ K_{39},\ K_{38}\}$ can be
reduced to $\{K_{8},\ K_{12},\ K_{38}\}$. Thus the number of the
incoherent Kraus operators can be reduced to 37.

Then, we discover that there is a unitary matrix $U_3$ which can
reduce the set $\{K_{4},\ K_{8},\ K_{38},\ K_{16},\ K_{12}\}$ to
$\{K_{4},\ K_{8},\ K_{12},\ K_{16}\}$. The specific form of $U_3$ is
as follows,
\begin{equation}
\begin{split}
 U_3=\left(\begin{array}{cccc}
       k_3a_4&0&k_3a_{38}&0\\
       -l_3a_{38}^2a_{16}|b_4|^2|b_{16}|^2&l_3a_8b_4b_{16}^*(a_4^2|b_{16}|^2+a_{38}^2|b_4|^2+a_{38}^2|b_{16}|^2)
       &l_3a_4a_{38}a_{16}|b_4|^2|b_{16}|^2&l_3a_{38}^2a_{16}|b_4|^3|b_{16}|\\
       m_3a_8a_{38}b_4^*b_{16}&-m_3a_{38}a_{16}|b_4|^2&m_3a_4a_8b_{16}b_4^*&ma_8a_{38}|b_4|^2\\
       n_3a_{38}^2b_4^*b_{16}&0&-n_3a_4a_{38}b_4^*b_{16}&n_3(a_4^2+a_{38}^2)|b_{16}|^2\\
       \end{array} \right),
 \end{split}
\end{equation}
where the parameters $k_3$, $l_3$, $m_3$ and $n_3$ are chosen as
\begin{equation}
  \begin{split}
  k_3^2&=\frac{1}{a_4^2+a_{38}^2},\\
  l_3^2&=\frac{1}{|b_4|^2|b_{16}|^2(a_{38}^4a_{16}^2|b_4|^2|b_{16}|^2+a_8^2(a_4^2|b_{16}|^2+a_{38}^2|b_4|^2+a_{38}^2|b_{16}|^2)^2
  +a_4^2a_{38}^2a_{16}^2|b_4|^2|b_{16}|^2+a_{38}^4a_{16}^2|b_4|^4)},\\
  m_3^2&=\frac{1}{|b_4|^2(a_8^2a_{38}^2|b_{16}|^2+a_{38}^4a_{16}^2|b_4|^2+a_4^2a_8^2|b_{16}|^2+a_8^2a_{38}^2|b_4|^2)},\\
  n_3^2&=\frac{1}{|b_{16}|^2(a_{38}^4|b_4|^2+a_4^2a_{38}^2|b_4|^2+(a_4^2+a_{38}^2)^2|b_{16}|^2)}.
  \end{split}
\end{equation}

Similarly, we can find some special unitary matrices reducing the set $\{K_{11},\ K_{12},\ K_{19},\ K_{20}\}$,
$\{K_{15},\ K_{16},\ K_{35},\ K_{36}\}$, $\{K_{15},\ K_{16},
\ K_{23},\ K_{24}\}$ and $\{K_{11},\ K_{12},\ K_{27},\ K_{28}\}$  to the set $\{K_{11},\ K_{12},\ K_{19}\}$,
$\{K_{15},\ K_{16},\ K_{35}\}$, $\{K_{15},\ K_{16},\ K_{23}\}$
and $\{K_{11},\ K_{12},\ K_{27}\}$ respectively. Therefore, the upper bound on the number of incoherent Kraus operators is 32.

\section{C: The proof of Theorem 2.}
In Ref. \cite{PhysRevLett.119.140402}, the authors verify that any
qutrits strictly incoherent operation admits a decomposition with at
most 15 incoherent Kraus operators. The specific form is as follows,

\begin{equation}
\begin{split}
&K_{1}= \left[ \begin{array}{ccc}
       a_{1}&0&0\\ 0&b_{1}&0\\0&0& c_{1}\\
      \end{array} \right],
\quad  K_{2}= \left[ \begin{array}{ccc}
       a_{2}&0&0\\0&0&c_{2}\\ 0&b_{2}&0\\
       \end{array} \right],
\quad  K_{3}= \left[ \begin{array}{ccc}
        0&b_{3}&0\\ a_{3}&0&0\\0&0&c_{3}\\
      \end{array} \right],
\quad  K_{4}= \left[ \begin{array}{ccc}
       0&0&c_{4}\\a_{4}&0&0\\ 0&b_{4}&0\\
      \end{array} \right],
\quad  K_{5}= \left[ \begin{array}{ccc}
       0&0&c_{5}\\ 0&b_{5}&0\\a_{5}&0&0\\
      \end{array} \right],\\
  &   K_{6}= \left[ \begin{array}{ccc}
        0&b_{6}&0\\0&0& c_{6} \\ a_{6}&0&0\\
      \end{array} \right],
\quad  K_{7}= \left[ \begin{array}{ccc}
       a_{7}&0&0\\ 0&b_{7}&0\\0&0&0\\
      \end{array} \right],
\quad  K_{8}= \left[ \begin{array}{ccc}
       a_{8}&0&0\\0&0&0\\ 0&b_{8}&0\\
      \end{array} \right],
\quad  K_{9}= \left[ \begin{array}{ccc}
        0&b_{9}&0\\ a_{9}&0&0\\0&0&0\\
      \end{array} \right],
\quad  K_{10}= \left[ \begin{array}{ccc}
       0&0&0\\a_{10}&0&0\\ 0&b_{10}&0\\
      \end{array} \right],\\
&     K_{11}= \left[ \begin{array}{ccc}
       0&0&0\\ 0&b_{11}&0\\a_{11}&0&0\\
      \end{array} \right],
\quad K_{12}= \left[ \begin{array}{ccc}
        0&b_{12}&0\\0&0&0 \\ a_{12}&0&0\\
      \end{array} \right],
\quad  K_{13}= \left[ \begin{array}{ccc}
       a_{13}&0&0\\ 0&0&0 \\0&0&0 \\
      \end{array} \right],
\quad  K_{14}= \left[ \begin{array}{ccc}
     0&0&0 \\ a_{14}&0&0\\0&0&0 \\
      \end{array} \right],
\quad  K_{15}= \left[ \begin{array}{ccc}
        0&0&0 \\0&0&0 \\ a_{15}&0&0\\
      \end{array} \right].
\end{split}
\end{equation}

By defining the following $3\times3$ unitary matrix
\begin{equation}
\begin{split}
 U_1=\left(\begin{array}{ccc}
 -l_1(|a_{12}|^2+|a_{15}|^2)b_9 & -l_1|a_{15}|^2b_{12}^\ast & l_1a_{12}a_{15}^\ast b_{12}^\ast\\
 0 & -m_1a_{12}^\ast & -m_1a_{15}^\ast\\
 n_1a_{15}b_{12} & -n_1a_{15}b_{9}^\ast &  n_1a_{12}b_{9}^\ast
       \end{array} \right),
 \end{split}
\end{equation}
where the parameters $l_1$, $m_1$ and $n_1$ are chosen as
\begin{equation}
  \begin{split}
  l_1^2&=\frac{1}{(|a_{12}|^2+|a_{15}|^2)(|a_{12}|^2|b_9|^2+|a_{15}|^2|b_9|^2
  +|a_{15}|^2|b_{12}|^2)},\\
  m_1^2&=\frac{1}{|a_{12}|^2+|a_{15}|^2},\\
  n_1^2&=\frac{1}{|a_{12}|^2|b_9|^2+|a_{15}|^2|b_9|^2+|a_{15}|^2|b_{12}|^2}.
  \end{split}
\end{equation}
we find the set $\{K_9, K_{12}, K_{14}, K_{15}\}$ can be reduced to
$\{K_9, K_{12}, K_{14}\}$.

Besides, we take the unitary matrix
\begin{equation}
\begin{split}
 U_2=\left(\begin{array}{ccc}
 -l_2(|a_{10}|^2+|a_{14}|^2)b_8 & -l_2|a_{14}|^2b_{10}^\ast & l_2a_{10}a_{14}^\ast b_{10}^\ast\\
 0 & -m_2a_{10}^\ast & -m_2a_{14}^\ast\\
 n_2a_{14}b_{10} & -n_2a_{14}b_{8}^\ast &  n_2a_{10}b_{8}^\ast
       \end{array} \right),
 \end{split}
\end{equation}
where the parameters $l_2$, $m_2$ and $n_2$ are chosen as
\begin{equation}
  \begin{split}
  l_2^2&=\frac{1}{(|a_{10}|^2+|a_{14}|^2)(|a_{10}|^2|b_8|^2+|a_{14}|^2|b_8|^2
  +|a_{14}|^2|b_{10}|^2)}\\
  m_2^2&=\frac{1}{|a_{10}|^2+|a_{14}|^2},\\
  n_2^2&=\frac{1}{|a_{10}|^2|b_8|^2+|a_{14}|^2|b_8|^2+|a_{14}|^2|b_{10}|^2}.
  \end{split}
\end{equation}
The combination of $K_8$, $K_{10}$ and $K_{14}$ has the same form of
$K_{13}$. So the set $\{K_8, K_{10}, K_{13}, K_{14}\}$ can be
reduced to $\{K_8, K_{10}, K_{13}\}$. In other words, any
single-qutrit strictly incoherent operation admits a decomposition
with at most 13 strictly incoherent Kraus operators.

\section{D: The computations of the conditions for the state transformation}

In this part, we will introduce the proof of the conditions for
the state transformation. For the 2-dimensional sections
$\sum_3(i,j)$, where $i\in\{1,2,...,6\}$ and $j\in\{7,8\}$, the
coefficients $m_i$ and $m_j$ of the final state via a strictly
incoherent channel can be derived as follows:
\begin{equation}
  \begin{split}
  m_1&=t_1(a_1Re[b_1]+a_3Re[b_3]+a_7Re[b_7]+a_9Re[b_9]),\\
  m_2&=t_2(a_1Re[c_1]+a_5Re[c_5]),\\
  m_3&=t_3(\frac{1}{2}(b_1c_1^\dagger+b_1^\dagger c_1+b_2c_2^\dagger+b_2^\dagger c_2)),\\
  m_4&=t_4(a_1Re[b_1]-a_3Re[b_3]+a_7Re[b_7]-a_9Re[b_9]),\\
  m_5&=t_5(a_1Re[c_1]-a_5Re[c_5]),\\
  m_6&=t_6(\frac{1}{2}(b_1c_1^\dagger+b_1^\dagger c_1-b_2c_2^\dagger+b_2^\dagger c_2)),\\
  m_7&=\frac{1}{3}(1-|c_1|^2-|c_3|^2)+(1-|a_5|^2-|a_6|^2-|a_{11}|^2-|a_{12}|^2)(\frac{1}{3}
  +\frac{t_7}{2})+(1-|b_2|^2-|b_4|^2-|b_{8}|^2-|b_{10}|^2)(\frac{1}{3}-\frac{t_7}{2}),\\
  m_8&=\frac{1}{\sqrt{3}}-\sqrt{3}((|c_1|^2+|c_3|^2)(\frac{1}{3}-\frac{t_8}{\sqrt{3}})
  +(|a_5|^2
  +|a_6|^2+|a_{11}|^2+|a_{12}|^2+|b_2|^2+|b_4|^2+|b_{8}|^2+|b_{10}|^2)(\frac{1}{3}
  +\frac{t_8}{2\sqrt{3}})).
  \end{split}
\end{equation}

Due to the completeness of Kraus operators, $\sum_iK_i^\dagger
K_i=\mathcal {I}$, we obtain
$\sum_{i=1}^{13}a_i^2=\sum_{j=1}^{12}|b_j|^2=\sum_{k=1}^{6}|c_k|^2=1$,
where $a_i$ can be chosen as real numbers, $b_j$ and $c_k$ as
complex numbers. It is easy to obtain the conditions 1 by using the
length of the Bloch vector, the normalization of the parameters and
the Cauchy¨CSchwarz inequality.

In the $m_7-m_8$ plane, we can obtain the explicit relation. The
final form of $m_7,\ m_8$ are as follows:
\begin{equation}
  \begin{split}
  m_7&=(|c_2|^2+|c_4|^2+|c_5|^2+|c_6|^2)(\frac{1}{3}-\frac{t_8}{\sqrt{3}})\\
&+(|b_1|^2+|b_3|^2+|b_5|^2
  +|b_6|^2+|b_7|^2+|b_9|^2+|b_{11}|^2+|b_{12}|^2)(\frac{1}{3}+\frac{1}{2}
  (-t_7+\frac{t_8}{\sqrt{3}}))\\
&+(a_1^2+a_2^2+a_3^2+a_4^2+a_7^2+a_8^2+a_9^2+a_{10}^2+a_{13}^2)(\frac{1}{3}
+\frac{1}{2}(t_7+\frac{t_8}{\sqrt{3}})),\\
 m_8&=\frac{1}{\sqrt{3}}-\sqrt{3}(|c_1|^2+|c_3|^2)(\frac{1}{3}-\frac{t_8}
 {\sqrt{3}})+(|b_2|^2+|b_4|^2+|b_8|^2
 +|b_{10}|^2)(\frac{1}{3}+\frac{1}{2}(-t_7+\frac{t_8}{\sqrt{3}}))\\
&+(a_5^2
 +a_6^2+a_{11}^2+a_{12}^2)
 (\frac{1}{3}+\frac{1}{2}(t_7+\frac{t_8}{\sqrt{3}})).
  \end{split}
\end{equation}

It is obvious that $-\sqrt{3}m_7+m_8-\frac{2\sqrt{3}}{3}=0$, using the conditions of $\sum_{i=1}^{13}a_i^2=\sum_{j=1}^{12}|b_j|^2
=\sum_{k=1}^{6}|c_k|^2=1$.

In $m_1-m_4$ plane, the final state form of $m_1,\ m_4$ are as
follows:
\begin{equation}
  \begin{split}
  m_1&=\frac{1}{2}((a_1b_1^\dagger+a_3b_3^\dagger+a_7b_7^\dagger+a_9b_9^\dagger)(t_1-it_4)
  +((a_1b_1+a_3b_3+a_7b_7+a_9b_9)(t_1+it_4)),\\
  m_4&=\frac{-i}{2}((-a_1b_1^\dagger+a_3b_3^\dagger-a_7b_7^\dagger+a_9b_9^\dagger)(t_1-it_4)
  +((a_1b_1-a_3b_3+a_7b_7-a_9b_9)(t_1+it_4)).\\
  \end{split}
\end{equation}
Then
\begin{equation}
 \begin{split}
 m_1^2+m_4^2&=(a_1^2|b_1|^2+a_3^2|b_3|^2+a_7^2|b_7|^2+a_9^2|b_9|^2+a_1a_7b_1b_7^\dagger
 +a_3a_9b_3b_9^\dagger+a_1a_7b_1^\dagger b_7+a_3a_9b_3^\dagger b_9)(t_1^2+t_4^2)\\
  &+(a_1a_3b_1^\dagger b_3^\dagger+a_3a_7b_3^\dagger b_7^\dagger+a_1a_9b_1^\dagger b_9^\dagger
  +a_7a_9b_7^\dagger b_9^\dagger+a_1a_3b_1 b_3+a_3a_7b_3 b_7+a_1a_9b_1 b_9+a_7a_9b_7 b_9)(t_1^2-t_4^2)
  \end{split}
\end{equation}

They are classified into 3 types of conditions:

$(1)$ When $t_1=t_4$, we get $m_1^2+m_4^2\leq t_1^2+t_4^2$ directly
by using the  Cauchy¨CSchwarz inequality.

$(2)$  When $(a_1a_3b_1^\dagger b_3^\dagger+a_3a_7b_3^\dagger b_7^\dagger+a_1a_9b_1^\dagger
b_9^\dagger+a_7a_9b_7^\dagger b_9^\dagger+a_1a_3b_1 b_3+a_3a_7b_3 b_7+a_1a_9b_1 b_9+a_7a_9b_7 b_9)(t_1^2-t_4^2)\leq0$,
 $m_1^2+m_4^2\leq t_1^2+t_4^2$ holds.

$(3)$ When $(a_1a_3b_1^\dagger b_3^\dagger+a_3a_7b_3^\dagger
b_7^\dagger+a_1a_9b_1^\dagger b_9^\dagger+a_7a_9b_7^\dagger
b_9^\dagger+a_1a_3b_1 b_3+a_3a_7b_3 b_7+a_1a_9b_1 b_9+a_7a_9b_7
b_9)(t_1^2-t_4^2)\geq0$, by setting $t_1> t_4 $, we find
\begin{equation}
\begin{split}
  m_1^2+m_4^2&=(a_1^2|b_1|^2+a_3^2|b_3|^2+a_7^2|b_7|^2+a_9^2|b_9|^2+a_1a_7b_1b_7^\dagger
  +a_3a_9b_3b_9^\dagger+a_1a_7b_1^\dagger b_7+a_3a_9b_3^\dagger b_9)(t_1^2+t_4^2)\\
  &+(a_1a_3b_1^\dagger b_3^\dagger+a_3a_7b_3^\dagger b_7^\dagger+a_1a_9b_1^\dagger b_9^\dagger
  +a_7a_9b_7^\dagger b_9^\dagger+a_1a_3b_1 b_3+a_3a_7b_3 b_7+a_1a_9b_1 b_9+a_7a_9b_7 b_9)(t_1^2-t_4^2)\\
 & \leq (a_1^2|b_1|^2+a_3^2|b_3|^2+a_7^2|b_7|^2+a_9^2|b_9|^2+2a_1a_7|b_1||b_7|+a_3a_9|b_3||b_9|)(t_1^2+t_4^2)\\
  &+2(a_1a_3|b_1||b_3|+a_3a_7|b_3||b_7|+a_1a_9|b_1||b_9|+a_7a_9|b_7||b_9|)(t_1^2-t_4^2)\\
  & \leq (a_1|b_1|+a_3|b_3|+a_7|b_7|+a_9|b_9|)^2 (t_1^2+t_4^2)\\
  & \leq (t_1^2+t_4^2)
  \end{split}
\end{equation}

Together with the three conditions, we show that the inequality
$m_1^2+m_4^2\leq t_1^2+t_4^2$ holds. Similar conditions can be
derived for $m_2-m_5$ plane and $m_3-m_6$ plane.

Without loss of generality, we take $m_1-m_2$ plane for the other
planes as an example. An IO maps a density matrix
$\{t_1,t_2,0,0,0,0,0,0\}$ to another density matrix
$\{m_1,m_2,0,0,0,0,0,0\}$, we have
\begin{equation}
  \begin{split}
  m_1&=(a_1Re[b_1]+a_3Re[b_3]+a_7Re[b_7]+a_9Re[b_9])t_1+((a_2Re[c_2]+a_4Re[c_4])t_1,\\
  m_2&=(a_2Re[b_2]+a_6Re[b_6]+a_8Re[b_8]+a_{12}Re[b_{12}])t_1+((a_1Re[c_1]+a_5Re[c_5])t_2.
  \end{split}
\end{equation}
Then, we find the relation between initial vector and the final
vector as
\begin{equation}
  \begin{split}
  (|m_1|+|m_2|)^2&=(|(a_1Re[b_1]+a_3Re[b_3]+a_7Re[b_7]+a_9Re[b_9]+a_2Re[b_2]+a_6Re[b_6]+a_8Re[b_8]+a_{12}Re[b_{12}])t_1|\\
             &+|(a_1Re[c_1]+a_2Re[c_2]+a_4Re[c_4]+a_5Re[c_5])t_2|)^2\\
             &\leq(|t_1+t_2|)^2.
  \end{split}
\end{equation}

Other 2-dimensional Bloch vectors have the similar relationship.


\smallskip
\begin{thebibliography}{99}
\bibitem{PhysRevLett.115.070503}  F. G. S. L. Brand\~ao and G. Gour,
Reversible Framework for Quantum Resource Theories, Phys. Rev. Lett.
115, 070503 (2015).
\bibitem{RevModPhys.91.025001} E. Chitambar and G. Gour, Quantum resource theories,
Rev. Mod. Phys. 91, 025001 (2019).
\bibitem{RevModPhys.81.865}R. Horodecki, P. Horodecki, M. Horodecki, K. Horodecki,
Quantum entanglement, Rev. Mod. Phys. 81, 865 (2009).
\bibitem{PhysRevLett.111.250404} F. G. S. L. Brand\~ao, M. Horodecki, J. Oppenheim,
J. M. Renes, R. W. Spekkens, Resource Theory of Quantum States Out of Thermal Equilibrium,
Phys. Rev. Lett. 111, 250404 (2013).
\bibitem{PhysRevLett.113.140401} T. Baumgratz, M. Cramer, and M. B. Plenio, Quantifying coherence,
Phys. Rev. Lett. 113, 140401 (2014).
\bibitem{Gyongyosi1}L. Gyongyosi, Correlation measure equivalence in dynamic causal structures of quantum gravity. Quantum Engineering. 2(1):e30(2020).
\bibitem{Gyongyosi2}L. Gyongyosi, and I. Sandor. Theory of quantum gravity information processing. Quantum Engineering 1(4):e23(2019).
\bibitem{humingliang}M.L. Hu, X.Y. Hu, J.C. Wang, Y. Peng, Y.R. Zhang, H. Fan, Quantum coherence and geometric quantum discord, Phys. Rep. 762, 1¨C100 (2018).
\bibitem{xi}Z.J. Xi,. Reverse coherent information and its properties. SCIENCE CHINA Physics, Mechanics $\&$ Astronomy 61,010321(2018).
\bibitem{Yang2018}L.M. Yang, B. Chen, S.M. Fei, and Z.X. Wang. Dynamics of coherence-induced state ordering under Markovian channels. Frontiers of Physics 13,130310(2018).
\bibitem{tao}M.J. Tao, N.N. Zhang, P.Y. Wen, F.G. Deng, Q. Ai, G.L. Long. Coherent and incoherent theories for photosynthetic energy transfer. Science Bulletin, 65, 318-328(2020).
\bibitem{Hillery2016Coherence}M. Hillery, Coherence as a resource in decision problems:
The Deutsch-Jozsa algorithm and a variation, Phys. Rev. A 93, 012111 (2016).
\bibitem{Lostaglio2011Description} M. Lostaglio, D. Jennings, and T. Rudolph, Description
of quantum coherence in thermodynamic processes requires constraints beyond free energy, Nat. Commun. 6, 6383 (2015).
\bibitem{Lostaglio2015Thermodynamic} M. Lostaglio, D. Jennings, and T. Rudolph, Thermodynamic
resource theories, non-commutativity and maximum entropy principles, New J. Phys. 19, 043008 (2017).
\bibitem{Micadei2015Coherent} K. Micadei, D. A. Rowlands, F. A. Pollock, L. C. C\'eleri, R. M. Serra,
and K. Modi, Coherent measurements in quantum metrology, New J. Phys. 17, 023057 (2015).
\bibitem{Lloyd2011Quantum} S. Lloyd, Quantum coherence in biological systems,
J. Phys.: Conf. Ser. 302, 012037 (2011).
\bibitem{PhysRevLett.116.120404} A. Winter and D. Yang, Operational resource theory of coherence,
Phys. Rev. Lett. 116, 120404 (2016).
\bibitem{PhysRevLett.119.140402} A. Streltsov, S. Rana, P. Boes, J. Eisert, Structure of the Resource
Theory of Quantum Coherence, Phys. Rev. Lett. 119, 140402 (2017).
\bibitem{Rana_2018} S. Rana and M. Lewenstein, Optimal decomposition of incoherent qubit channel,
J. Phys. A: Math. Theor. 51, 414002 (2018).
\bibitem{Nielsen2010Quantumbibtex} M. A. Nielsen and I. L. Chuang, Quantum Computation and Quantum
Information, 10th ed. (Cambridge University Press,2010).

\bibitem{KIMURA2003339}G. Kimura, The Bloch Vector for N-Level Systems, Phys. Lett. A 314, 339(2003).
\bibitem{JAKOBCZYK2001383} L. Jak{\'o}bczyk, M. Siennicki, Geometry of Bloch vectors
in two-qubit system, Phys. Lett. A 286, 383(2001).
\bibitem{PhysRevLett.119.230401} T. Theurer, N. Killoran, D. Egloff, and M. B. Plenio, Resource Theory of Superposition, Phys. Rev. Lett. 119,230401(2017).

\end{thebibliography}
\end{document}